\documentclass[journal]{IEEEtran}
\usepackage{amsmath,amssymb,amsfonts,amsthm}
\usepackage{amsmath}
\usepackage[caption=false,font=footnotesize,labelfont=rm,textfont=rm]{subfig}
\usepackage{graphicx}
\usepackage[dvipsnames, svgnames, x11names]{xcolor}
\usepackage{bbm}

\newcommand{\red}[1]{\color{red}#1}
\usepackage[colorlinks=true,linkcolor=red,citecolor=green]{hyperref}

\usepackage{algorithmic,algorithm}
\usepackage{setspace}

\usepackage{float}
\usepackage{subfig}
\usepackage{overpic}
\usepackage{epstopdf}
\usepackage{cases}
\usepackage{bm}

\usepackage{stfloats} 

\allowdisplaybreaks[4]

\usepackage{makecell} % 在导言区添加这行
\usepackage{booktabs}
\usepackage{multirow}

\begin{document}
	\title{Towards World Model–Empowered Integrated Sensing, Communication, and Decision for Complex Unmanned Systems}
	
	\author{
		\IEEEauthorblockN{Xue Han,  Yongpeng Wu, Meng Shen, Wenjun Xu, Biqian Feng, Zijin Wang, \\ Xiaohu You, Shengli Sun, and Wenjun Zhang}
		
	\thanks{Xue Han, Yongpeng Wu, Biqian Feng, Zijin Wang, and Wenjun Zhang are with the School of Integrated Circuits (School of Information Science and Electronic Engineering), Shanghai Jiao Tong University, Shanghai 200240, China (e-mail: han.xue@sjtu.edu.cn;  yongpeng.wu@sjtu.edu.cn; 
		fengbiqian@sjtu.edu.cn;
		zijin\_wang@sjtu.edu.cn;  zhangwenjun@sjtu.edu.cn) }
		
	\thanks{Meng Shen is with the Pervasive Communication Research Center, Purple Mountain Laboratories, Nanjing 211111, China (e-mail: shenmeng@pmlabs.com.cn).}

	\thanks{Wenjun Xu and Shengli Sun are with Shanghai Institute of Technical Physics, Chinese Academy of Sciences, Shanghai 200083, China (e-mail: xuwenjun@mail.sitp.ac.cn; Palm\_sun@mail.sitp.ac.cn).}
		
	\thanks{Xiaohu You is with the National Mobile Communications Research Laboratory, Southeast University, Nanjing 210096, China, and also with the Purple Mountain Laboratories, Nanjing 211111, China (e-mail: xhyu@seu.edu.cn).
		(Corresponding author:  Yongpeng Wu and Meng Shen).}
}

	\maketitle
\begin{abstract}
	Complex unmanned systems comprising satellites, unmanned aerial vehicles (UAVs), unmanned ground vehicles (UGVs), and quadruped robots are increasingly deployed to perform large-scale sensing and autonomous operations. 
	We propose a world model–empowered sensing, communication, decision (SCD) integration framework for complex unmanned communication networks. 
	The proposed architecture establishes a closed-loop system where a unified world model jointly optimizes time-sensitive sensing, wireless communication, and intelligent decision-making. 
	To regulate sensing freshness and reduce redundant data generation, we propose a time-sensitive age of information (AoI)-driven sensing mechanism that dynamically schedules sensing updates based on task urgency and predictive uncertainty. 
	Furthermore, a predictive world model is developed to jointly represent environmental dynamics, wireless channel evolution, and agent mobility within a hybrid deterministic–stochastic latent space. This enables proactive communication scheduling and decision evaluation via latent rollout. 
	To support large-scale heterogeneous coordination, a multi-granularity knowledge graph is further designed to organize cross-population relationships among satellites, UAVs, UGVs, and ground agents. 
Numerical results demonstrate that the proposed SCD framework outperforms conventional systems, highlighting the significant potential of world models for supporting unmanned systems.
\end{abstract}

%\vspace{-0.2cm}	
	\section{Introduction}
The development of complex unmanned systems, including satellites, unmanned aerial vehicles (UAVs), unmanned ground vehicles (UGVs), and quadruped robots, is transforming autonomous platforms from isolated agents into coordinated large-scale systems. As these heterogeneous systems are deployed in highly dynamic and partially observable environments, their effectiveness critically depends on their ability to perceive surroundings, share information, and execute collaborative actions \cite{plan_2024}. Traditionally, sensing, communication, and action processes have been treated as a modular, sequential pipelines. Sensing modules extract features, communication links transmit compressed packets, and decision modules react to the received data. While analytically tractable, this reactive and bit-centric separation exhibits inherent fragility in practice. It struggles to cope with the stringent latency demands, severe channel fluctuations, and long-horizon coordination required by next-generation unmanned swarms \cite{ding2025understanding}.

%The rapid evolution of artificial intelligence (AI) is shifting from task specific learning toward general-purpose intelligent systems that can perceive, reason, and act in complex environments. Central to this transition is the concept of world models, which aim to construct internal representations of external environments and predict their future evolution, enabling agents to perform long horizon planning and decision making through imagination rather than exhaustive real world interaction \cite{plan_2024}. A recent comprehensive survey formally characterizes world models as serving two fundamental purposes: understanding the mechanisms of the world via implicit representations, and predicting future states to support simulation and control \cite{ding2025understanding}.

To address these limitations, world models have emerged as a key paradigm for learning latent representations of environment dynamics and enabling long-horizon prediction through imagination. Originating in cognitive science and popularized in model-based reinforcement learning (MBRL) \cite{MBRL} by frameworks like Dreamer \cite{hafner2019dream}, world models empower agents to construct compact, internal representations of external environments and predict their future evolution. %Rather than relying on exhaustive real-world trial and error, agents can perform long-horizon planning through latent imagination. 
%However, purely statistical latent models often suffer from compounding errors and limited long-term coherence. 
Recent breakthroughs in generative artificial intelligence (AI) have further scaled these capabilities. Models such as the Dual-Mind World Model (DMWM) \cite{wang2025dmwm} integrate logic-guided reasoning, while 3D-enhanced architectures like Point World \cite{huang2026pointworld} enable zero-shot physical manipulation. 
%Rather than operating in pixel or latent vector spaces, Point World unifies states and actions as 3D point flows, predicting full-scene geometric evolution conditioned on robot actions. 
Simultaneously, frameworks like Local-to-Global (LOGO) \cite{liPuzzleItOut2026} have demonstrated how local agent dynamics can be aggregated into global transition models for multi-agent coordination. 
TD-MPC2 \cite{TDMPC} derives implicit world models, enabling agents to perform model predictive control directly in the latent space.

Beyond robotics and multi-agent control, world models are increasingly explored in communication and cyber-physical systems, where anticipating future network states is essential for resource allocation and quality-of-service optimization \cite{robot}. 
In autonomous driving, for instance, world models process real-time environmental data to perceive road conditions \cite{Driving} and forecast complex evolutionary trends \cite{Driving_CVPR}, establishing a foundation for immediate situational awareness.
Recent studies employ learned world models to predict network dynamics and guide scheduling decisions. This enables agents to minimize metrics like Age of Information (AoI) through imagined rollouts rather than continuous online interaction \cite{wang2025world}. 
%These efforts reveal the broader relevance of world models as a unifying abstraction for sequential decision-making across embodied, multi-agent, and communication domains.

%Despite these remarkable advancements, the application of world models remains fragmented across isolated domains of robotics, vision, and single-agent control. A critical gap persists in extending these capabilities to networked, multi-agent communication systems. In complex unmanned networks, dynamics are high-dimensional and strongly coupled; a UAV's mobility decision immediately alters the spatial topology, which in turn dictates wireless channel quality and the freshness of shared sensory data. Without a unified cognitive architecture that co-models physical movement and communication dynamics, current systems cannot fully transition from reactive control to proactive, uncertainty-aware reasoning.

Despite these advances, existing world model applications remain largely fragmented across robotics, vision, and single-agent control. A key limitation is the lack of a unified modeling framework for networked multi-agent unmanned systems, where mobility, wireless channels, and information freshness are tightly coupled. In such systems, physical movement directly reshapes network topology, which in turn affects communication quality and semantic information timeliness, making purely reactive or decoupled designs insufficient for coordinated decision-making.

To bridge this gap, this article takes the first step toward future scenarios
and proposes a novel world model-empowered integrated sensing, communication, and decision (SCD) framework for complex unmanned systems. By elevating the world model from a task-specific predictor to a system-wide cognitive infrastructure, the SCD pipeline evolves into a tightly coupled and predictive closed-loop framework. In this framework, sensing is driven by time-sensitive AoI metrics, communication is proactively orchestrated based on anticipated spatiotemporal network states, and decision-making is guided by a dynamic, multi-granularity knowledge graph (MGKG).
{{Unlike conventional model-based reinforcement learning (MBRL) that relies on black-box state transitions, our technical contribution lies in structurally embedding physical-layer communication dynamics and AoI evolution into the latent space, while leveraging an MGKG to bridge isolated world models for cross-population coordination.}}
The main contributions of this article can be summarized as follows.

\begin{figure*}[t]
	%\vspace{-0.33cm}  %调整图片与上文的垂直距离
	\setlength{\abovecaptionskip}{0.2cm}   %调整图片标题与图距离
	\setlength{\belowcaptionskip}{-3cm}   %调整图片标题与下文距离
	\centering
	\includegraphics[width=0.869\linewidth]{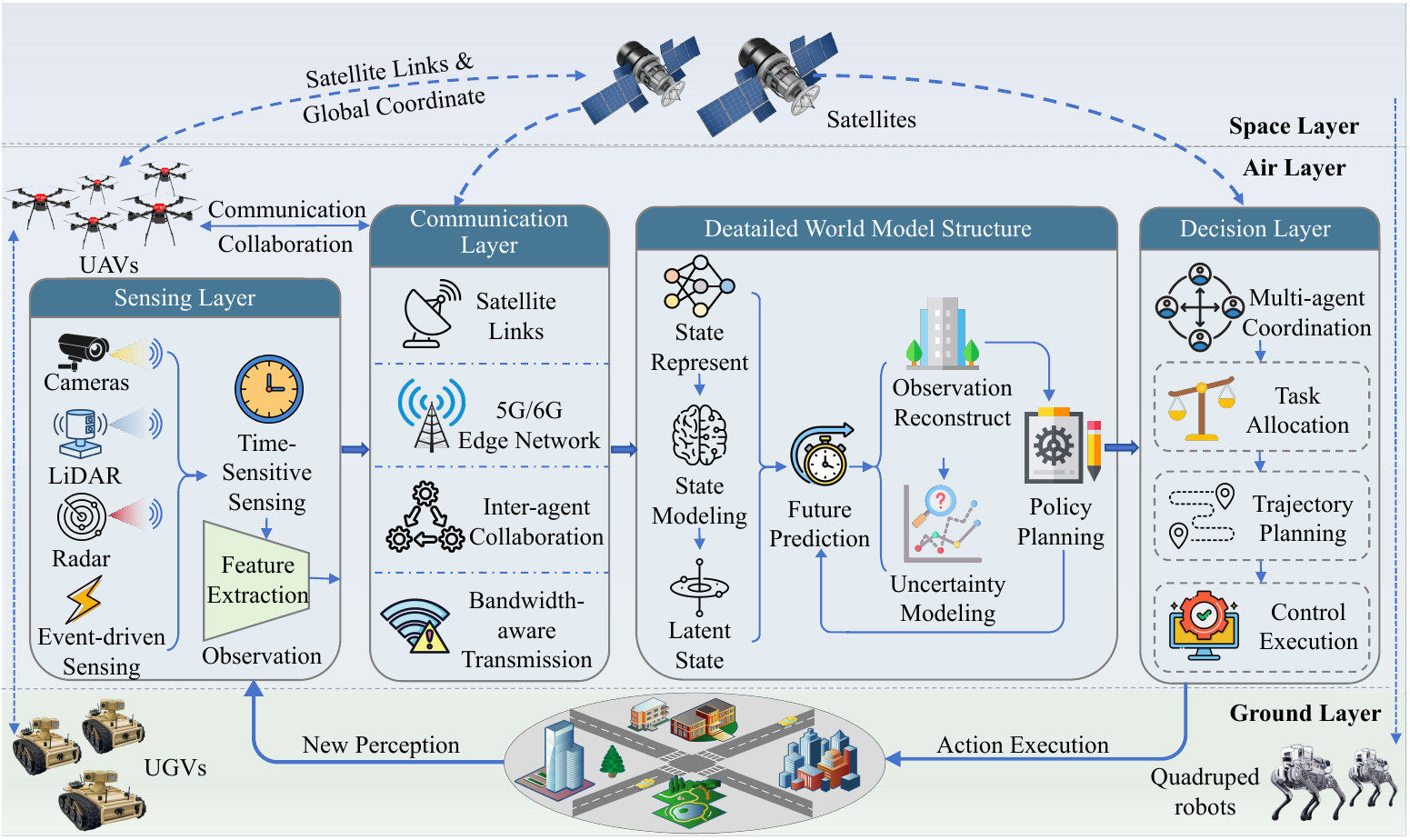}
	\caption{World model-empowered integrated sensing, communication, and decision framework for the complex and heterogeneous unmanned systems.}
	\label{system}
	\vspace{-0.51cm}
\end{figure*}

\begin{itemize}
\item {\textit{World Model–empowered SCD Integration Framework:} 
	We propose a world model–empowered  SCD integration architecture for complex unmanned networks, enabling closed-loop coordination across heterogeneous space-air-ground agents.}

\item {\textit{Time-Sensitive AoI-Driven Sensing:} 
	We design an AoI-aware sensing mechanism that reduces redundant data acquisition through event-triggered and semantic-aware sensing, improving communication efficiency while preserving situational awareness.}

\item {\textit{Predictive World Model and Cross-Layer Coordination:} 
	We develop a predictive world model that jointly captures environmental dynamics, mobility, and wireless channel evolution, enabling proactive communication scheduling and cooperative decision-making. We further design a MGKG to support cross-agent coordination.}

{\item \emph{Performance Validation:} 
Numerical results demonstrate significant improvements over conventional modular and reinforcement learning baselines.
}

\end{itemize}

 \begin{figure*}[t]
	\vspace{-0.33cm}  %调整图片与上文的垂直距离
	\setlength{\abovecaptionskip}{0.2cm}   %调整图片标题与图距离
	\setlength{\belowcaptionskip}{-3cm}   %调整图片标题与下文距离
	\centering
	\includegraphics[width=0.89\linewidth]{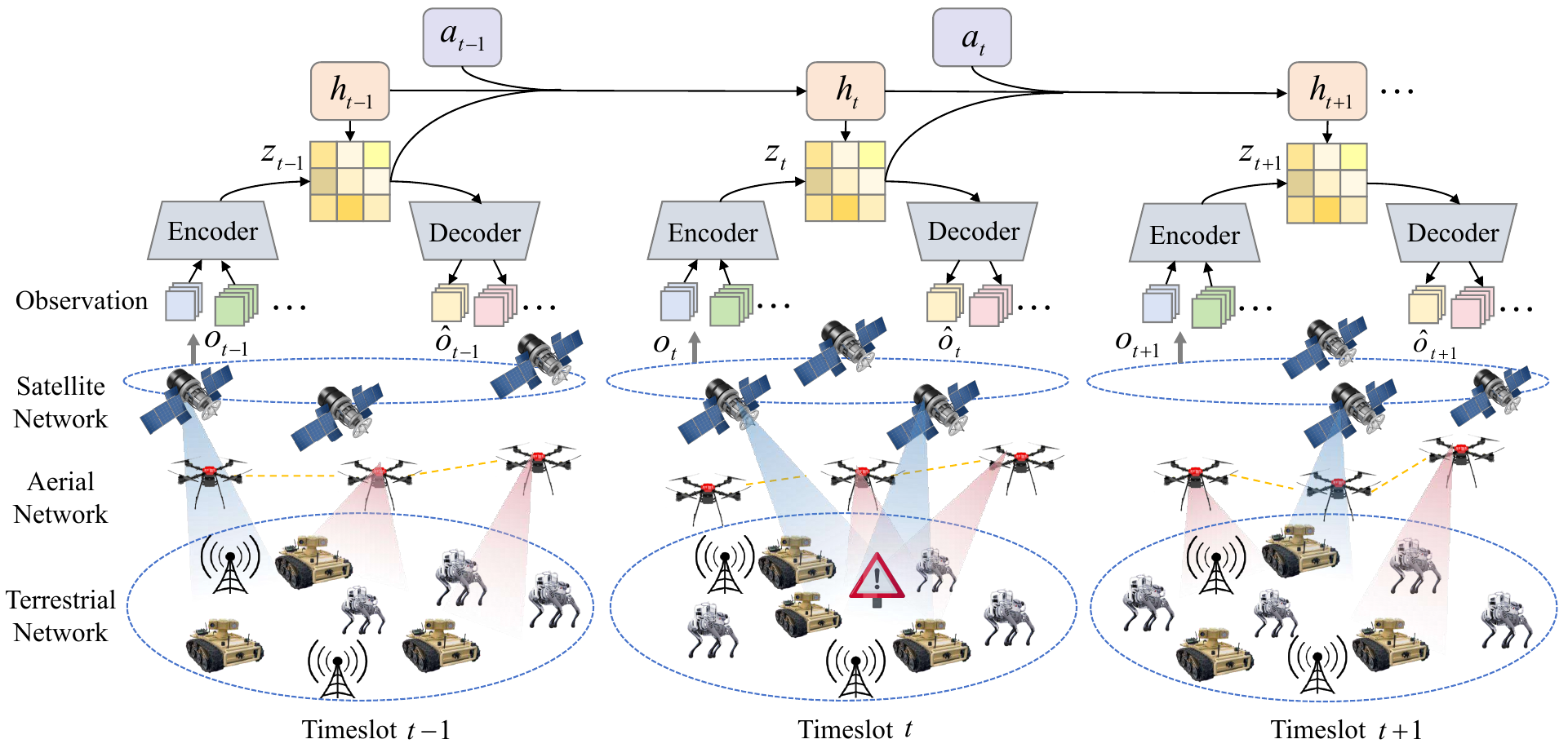}
	\caption{Learning a world model for complex unmanned communication networks based on the observed data.}
	\label{updating}
	\vspace{-0.51cm}
\end{figure*}

	\section{Why World Models for Complex Unmanned Systems}

World models have recently demonstrated broad applicability across diverse intelligent systems. 
%As illustrated in Fig. \ref{application}, world model-empowered frameworks significantly enhance multiple emerging application domains.
By learning compact latent dynamics and enabling counterfactual reasoning, world models convert reactive systems into predictive, coordinated, and uncertainty-aware intelligent infrastructures. Motivated by this paradigm shift, we investigate how world models can fundamentally reshape sensing, communication, and decision in complex unmanned systems.

Complex unmanned systems, such as satellites, UAVs, UGVs, and quadruped robots are evolving from isolated autonomous agents to highly coordinated, large-scale cyber-physical collectives \cite{AirScape}. These systems operate in dynamic, partially observable environments, where decision quality critically depends on the ability to reason over long temporal horizons and uncertain observations.

Traditional sensing, communication, and decision pipelines are typically designed in a modular manner. While such designs have enabled significant progress, they exhibit several inherent limitations in complex and rapidly changing environments. 
{First}, sensing modules often focus on instantaneous or short-term observations, with limited capability to explicitly capture long-horizon environmental dynamics. 
{Second}, communication mechanisms are primarily optimized for reliable data delivery, without explicitly accounting for how transmitted information contributes to downstream inference or control. 
{Third}, decision is commonly performed based on locally available information or partially shared states, which may lead to inconsistencies in environment understanding across distributed agents.

World models provide a fundamentally different paradigm. 
{{Unlike digital twins that passively mirror physical environments, predictive networking that primarily forecasts communication states, semantic communication that focuses on efficient information delivery, and conventional model-based reinforcement learning (RL) that optimizes control policies, the proposed framework establishes a unified world model as a cognitive infrastructure that jointly predicts, coordinates, and optimizes sensing, communication, and decision processes across heterogeneous unmanned systems.}}
For complex unmanned systems, world models offer three key advantages:
i) \emph{Predictive modeling:} Agents can anticipate future environmental states by leveraging learned latent dynamics, enabling informed decision-making over extended time horizons rather than relying solely on instantaneous observations. 
ii) \emph{Compact state representation:} High-dimensional sensory inputs are encoded into structured latent representations that capture essential environmental features, improving the efficiency of information processing and exchange. 
iii) \emph{Consistent environment understanding:} Shared or aligned latent beliefs allow distributed agents to maintain a coherent and continuously updated representation of the environment, facilitating coordinated decision-making.
Overall, world models provide a systematic framework for bridging perception, communication, and control in complex unmanned systems. By incorporating explicit environment modeling and predictive inference, they enable more robust, adaptive, and coordinated system behavior in dynamic and uncertain operating conditions.

\vspace{-0.2cm}
	\section{An Overview of World Model–Empowered SCD Integration Framework}
In this section, we introduce the detailed structure of the world model-empowered SCD system.
The proposed SCD integration framework establishes a closed-loop architecture that tightly couples time-sensitive sensing, adaptive wireless communication, structured world modeling, and intelligent decision-making. 
As illustrated in Fig. \ref{system}, heterogeneous sensing agents deployed across space, air, and ground layers collaboratively acquire global situational information. The communication layer dynamically coordinates multi-domain information exchange, while the world model serves as the cognitive core that unifies state representation, uncertainty modeling, and future prediction. Finally, the decision layer performs task allocation, trajectory planning, and multi-agent coordination and continuously refines its policy through knowledge graph updates.

\vspace{-0.2cm}
\subsection{Time-Sensitive Sensing} 
In these complex unmanned systems, sensing is no longer an isolated sensing task but a time-critical information acquisition process tightly coupled with communication and decision-making.
The sensing layer is responsible for multi-layer, cross-domain sensing to achieve full-domain situational awareness. Diverse sensing modalities including cameras, LiDAR, radar, and event-driven sensors, are deployed across satellites, UAVs, and UGVs to capture complementary environmental features. 
The emergence of large-scale low Earth orbit (LEO) satellite constellations, such as Starlink, provides a new infrastructure foundation for constellation-level time-sensitive sensing.

\begin{table*}[t] 
	\centering
	\renewcommand{\arraystretch}{1.5}
	\caption{Comparison Between Conventional Learning and World Model-Empowered Learning}
	\label{tab:worldmodel_comparison}
	
	% 调整为5列，重新分配宽度，减少换行带来的垂直空间占用
	\begin{tabular}{>{\centering\arraybackslash}m{3cm} m{3.1cm} m{3.2cm} m{3.35cm} m{3cm}}
		\toprule
		\textbf{Aspect} & 
		\multicolumn{1}{c}{\textbf{Core Paradigm}} & 
		\multicolumn{1}{c}{\textbf{Data Source}} & 
		\multicolumn{1}{c}{\textbf{Channel Modeling}} & 
		\multicolumn{1}{c}{\textbf{System Adaptation}} \\
		\midrule
		
		\textbf{\begin{tabular}[c]{@{}c@{}}Conventional \\ Learning\end{tabular}}
		& Unified model aiming for broad generalization. 
		& Large-scale and heterogeneous datasets. 
		& Statistical properties based on the historical observations. 
		& Reactive adaptation; very vulnerable to distribution shifts. \\
		\hline
		
		\textbf{\begin{tabular}[c]{@{}c@{}}World Model-\\ Empowered Learning\end{tabular}}
		& Environment-specific modeling tailored to local scenarios. 
		& Localized data (e.g., real-time sensing, digital twins). 
		& Environment-aware spatial geometry and propagation structures. 
		& Proactive optimization via learned predictive dynamics. \\
		
		\bottomrule
	\end{tabular}
	\vspace{-0.28cm}
\end{table*}

Within a world model–empowered SCD architecture, time-sensitive sensing plays a central role in regulating information freshness. 
To address the stringent latency and reliability requirements of complex unmanned systems, we propose a time-sensitive AoI sensing mechanism to establish a novel information freshness assessing paradigm \cite{AoI}.  
{{Unlike conventional AoI-aware scheduling that mainly optimizes update transmission, the proposed mechanism proactively activates sensing according to task urgency, environmental dynamics, and predicted state uncertainty.
Furthermore, edge-based feature extraction enables semantic-level abstraction of raw sensory data, allowing sensing and communication to be jointly coordinated within the closed-loop SCD framework. The sensing layer thus forms a structured and temporally aligned observation stream for the world model.}}

For time-critical decision-making, the key metric is not merely throughput but the time sensitive AoI, which can be expressed as the time difference between the current time and the generation time of the most recently received update. A large AoI implies outdated state knowledge, leading to inaccurate environment rollout and degraded policy evaluation. By leveraging inter-satellite links and distributed relay scheduling, constellation-based time-sensitive sensing minimizes AoI through fast in-orbit fusion and adaptive forwarding. More importantly, the world model itself can guide sensing frequency based on its internal uncertainty estimation. This establishes a feedback loop in which sensing and cognition co-evolve, transforming sensing from a passive acquisition process into an uncertainty-aware adaptive mechanism.

In large-scale unmanned swarms, this leads to a hierarchical closed-loop process: distributed sensing generates fresh observations, communication networks deliver structured updates, the world model performs state estimation and prediction, and decision modules adjust policies accordingly. Continuous observation injection prevents the architecture from degenerating into an open-loop predictor and ensures robust adaptation under dynamic conditions.

\vspace{-0.32cm}
\subsection{Wireless Communication}
In conventional wireless systems, the communication layer is largely built upon statistical channel abstractions, such as Rayleigh or Rician fading models, where the physical world is represented through probabilistic approximations. While analytically tractable, such models are fundamentally coarse-grained and reactive.
They describe average channel behavior but lack structural awareness of the environment, mobility dynamics, and task semantics. In complex unmanned systems, where satellites nodes, UAVs, UGVs, and quadruped robots operate across space-air–ground domains, these statistical models become insufficient. High mobility, intermittent blockage, heterogeneous traffic (e.g., real-time video streaming versus low-rate control signaling), and rapidly evolving topology create communication conditions that cannot be effectively captured by static probabilistic assumptions alone. As a result, traditional sensing and decision execution chains become fragile when the communication link fluctuates.

Table \ref{tab:worldmodel_comparison} contrasts the proposed world model-driven architecture with conventional learning approaches.
The communication layer in a world model–empowered SCD framework exhibits a fundamentally different design philosophy. Rather than treating the wireless channel as an uncertain random process requiring passive adaptation, the system embeds communication dynamics directly into the world model. The world model maintains a structured representation of environmental geometry, agent mobility, traffic demand, and even network topology evolution. Consequently, channel quality is no longer inferred solely from short-term signal measurements but predicted based on spatial configuration, trajectory rollout, and anticipated interference patterns. This transforms communication from reactive adaptation to predictive orchestration. Beam alignment, relay selection, spectrum allocation, and handover decisions can be proactively scheduled according to future network states inferred by the world model, significantly improving robustness under high dynamics.

 \vspace{-0.2cm}
\subsection{Detailed World Model Updating Mechanism} 
The world model in a complex unmanned system serves as a unified latent representation of the physical environment, network conditions, agent states, and mission objectives. Unlike conventional situational awareness modules that provide instantaneous snapshots, the world model maintains a temporally evolving internal state that integrates sensing inputs, communication feedback, and decision outcomes. This persistent internal memory enables predictive reasoning and coordinated multi-agent adaptation under highly dynamic and partially observable environments.

\subsubsection{Observation and State Representation}
At time slot $t$, the world model receives multimodal observations,
$o[t]=\{A[t], P[t], L[t]\}$, where $A[t]$ denotes the time-sensitive AoI of each agent, $P[t]$ represents sensing-related information, and $L[t]$ denotes agent locations.
Specifically, the time-sensitive AoI allows the system to quantify the freshness and reliability of distributed information, which is critical for cooperative sensing and coordinated control. The location information provides geometric context, enabling the model to infer spatial relationships among transmitters and receivers, including relative distance, orientation, and potential line-of-sight (LoS) conditions.
The sensing data further reflects instantaneous wireless propagation characteristics, such as multipath effects, dynamic blockage events, and Doppler shifts, which are particularly prominent in high-frequency bands and highly mobile scenarios.
{{These heterogeneous observations are encoded into a latent space designed to align different temporal scales, composed of: i) a deterministic hidden state $h[t]$, capturing slow-scale macroscopic dynamics such as AoI evolution, agent mobility, spatial topology and mission status; and ii) a stochastic latent variable $z[t]$, modeling the aggregated distributional uncertainty of fast-scale microscopic processes like wireless channel variations and interference.}} By synchronizing probabilistic fast-scale features with deterministic slow-scale states at each timeslot, this hybrid representation allows the world model to jointly learn predictable evolution patterns and residual uncertainties.

 \begin{figure}[t]
	\vspace{-0.33cm}  %调整图片与上文的垂直距离
	\setlength{\abovecaptionskip}{-0.12cm}   %调整图片标题与图距离
	\setlength{\belowcaptionskip}{-3cm}   %调整图片标题与下文距离
	\centering
	\includegraphics[width=1\linewidth]{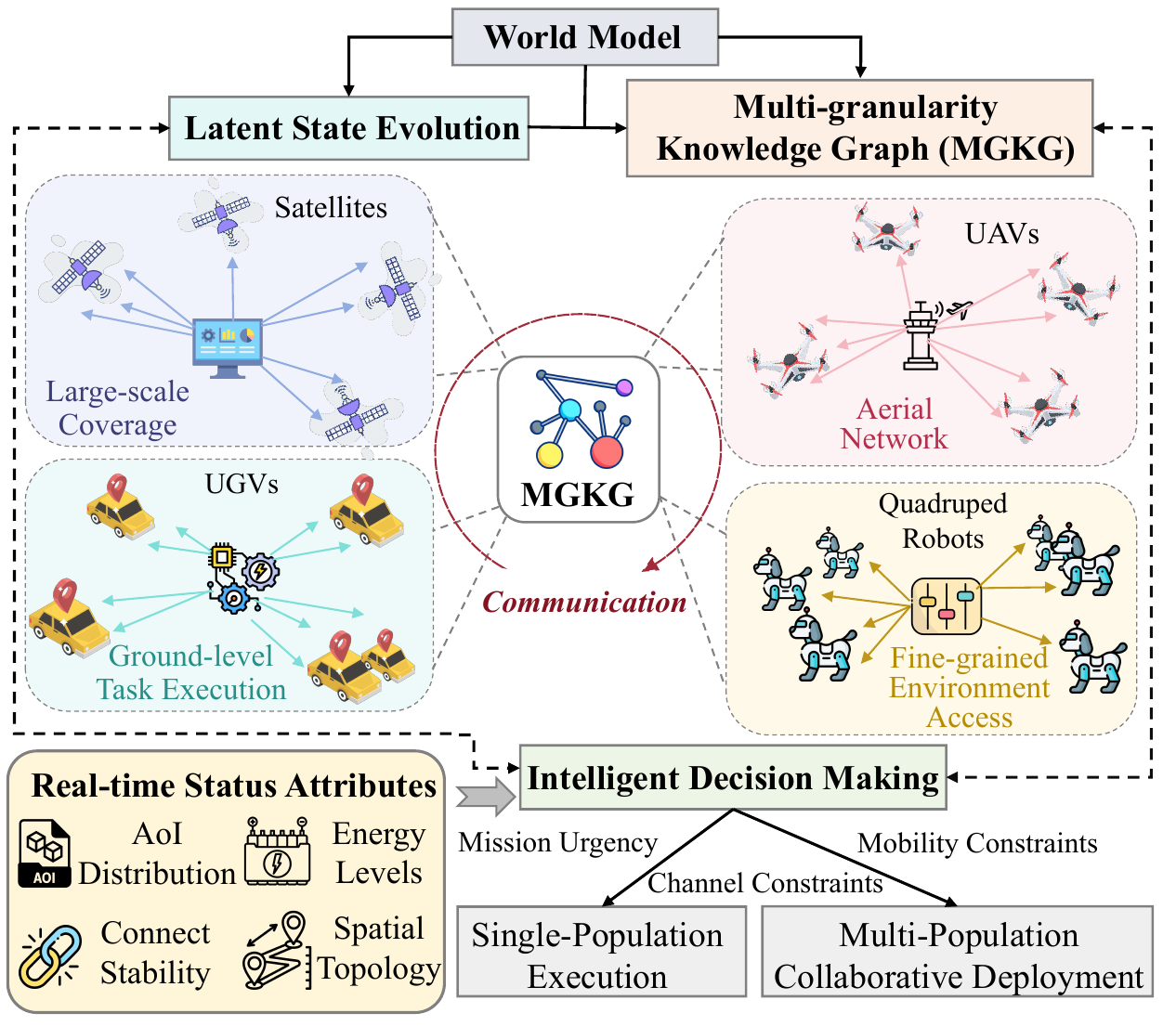}
	\caption{The multi-granularity knowledge graph (MGKG) structure.}
	\label{knowledgeG}
	\vspace{-0.31cm}
\end{figure}

\subsubsection{Latent State Transition and Predictive Rollout}
{{As illustrated in Fig. \ref{updating}, the state transition of world model follows
$(h[t],z[t])=f_\mathrm{trans}(h[t-1],z[t-1],a[t-1])$,
where the action vector $a[t-1]$ comprising physical-layer optimization iterates (e.g., power allocation). These iterates serve as explicit conditional inputs to $f_\mathrm{trans}$, directly modulating the distribution parameters of the future latent features $(h[t], z[t])$.
The latent transition function jointly captures the coupled evolution of information freshness, agent mobility, wireless propagation, and communication actions. Physical movement alters network topology and channel conditions, communication scheduling affects AoI evolution, and newly acquired observations continuously update the latent belief, resulting in an action-conditioned closed-loop system dynamics model.
Therefore, the world model captures not only environmental dynamics but also action-conditioned system evolution.}}
Furthermore, the learned transition function supports predictive rollout in latent space. Given candidate actions $\{a[t], a[t+1], \dots\}$, the model can simulate future trajectories $(h[t + \tau],z[t + \tau])$, without interacting with the physical system. This mechanism enables proactive link scheduling and cooperative strategy evaluation, significantly reducing real-world trial-and-error overhead.

\subsubsection{Reconstruction and Consistency Learning}
To ensure representation fidelity, the latent state is decoded to reconstruct observations: $\hat o[t]=f_\mathrm{dec}(h[t],z[t])$. The reconstruction objective enforces consistency between predicted and actual AoI evolution, spatial topology, and propagation characteristics. In addition, temporal consistency constraints are imposed to preserve smooth mobility patterns and realistic channel transitions.

\subsubsection{System-Level Coordination Mechanism}
A distinctive feature of complex unmanned systems lies in their tightly coupled SCD loop. The world model serves as the coordination core that synchronizes: i) AoI-aware sensing: Triggering new observations when the weighted sum of current AoI and the predictive variance of the latent state $z[t]$ exceeds a threshold.
ii) Predictive communication scheduling: The predicted channel evolution is used to determine relay selection, beam alignment, bandwidth allocation, and transmission scheduling.
iii) Cooperative decision-making: Candidate actions are evaluated via latent rollout before physical execution, allowing the system to select policies with lower expected uncertainty and communication cost.
{{The world model unifies SCD interactions in a tight loop: sensing freshness drives communication scheduling, network latency constrains decision reliability, and mobility decisions reshape the spatial topology to physically alter future channels and sensing coverage.}}

{{\subsubsection{Unique Mechanisms for Complex Unmanned Systems}
Compared with traditional digital twins or network state estimators, the proposed world model acts as an active, action-conditioned cognitive engine, distinguished by:
i) Action-aware channel evolution modeling: Communication decisions directly influence future state transitions.
ii) AoI-driven temporal prioritization: Information freshness is embedded into the state space as a first-class dynamic variable.
iii) Joint spatial–temporal reasoning: Mobility geometry and wireless propagation are co-modeled via a unified latent representation.
iv) Predictive cooperative simulation: Multi-agent coordination is evaluated via latent rollouts before real deployment.
v) Cross-layer integration: Physical-layer channel dynamics, medium access control (MAC)-layer scheduling, and system-level mission objectives are jointly represented.
These mechanisms enable the world model to function not merely as a predictor of wireless channels, but as a cognitive infrastructure that orchestrates unmanned agents under stringent latency, reliability, and adaptability requirements.}}

\vspace{-0.2cm}
\subsection{Intelligent Decision and Knowledge Graph Update}
Beyond latent state evolution, the world model further maintains a structured knowledge layer that explicitly organizes accumulated system experience into a dynamic knowledge graph (KG) \cite{KG}. While the latent variables $(h[t], z[t])$ capture compact representations for prediction and control, the knowledge graph provides interpretable, multi-granular, and relational reasoning capabilities.

In Fig. \ref{knowledgeG}, the MGKG emphasizes cross-population construction among heterogeneous unmanned clusters, such as satellites, UAVs, UGVs, and quadruped robotic platforms.
In complex missions, these platforms do not operate as isolated agents but as functional populations with complementary sensing ranges, mobility constraints, energy budgets, and communication capabilities. The KG therefore encodes not only intra-group relationships but more critically, inter-group semantic and functional dependencies.
For instance, satellite nodes may provide large-scale coverage and backhaul connectivity. UAVs offer agile aerial relaying and dynamic sensing. Unmanned vehicles ensure ground-level task execution. Quadruped robots perform fine-grained inspections in obstructed or confined environments.
Edges in the graph represent capability complementarity, communication reachability, task coupling, and relay feasibility across these distinct populations.

Importantly, this cross-population knowledge graph supports flexible dispatching and joint operation mechanisms. When mission urgency, channel degradation, or mobility constraints arise, the world model can query the graph to dynamically reconfigure cooperation patterns, activating single-population execution or multi-population collaborative deployment. The graph encodes not only static capabilities but also real-time status attributes such as AoI distribution, energy levels, connectivity stability, and spatial topology, allowing coordinated decision-making that balances communication efficiency and mission effectiveness.

Through continuous updating from latent state predictions and observed outcomes, the inter-population knowledge graph evolves to capture recurring cooperation patterns and environmental regularities.
{{Crucially, these abstracted regularities serve as domain-invariant prior knowledge. When deployed in unseen environments, the MGKG provides a graph-based prior to accelerate few-shot adaptation, allowing the local world model to rapidly calibrate without extensive retraining.}}
This mechanism transforms heterogeneous unmanned clusters into an adaptive, orchestrated system-of-systems, where communication, sensing, and action are jointly optimized across domains rather than confined within individual platforms.
{{For practical deployment in large-scale unmanned systems, the framework can leverage a distributed edge computing architecture, where resource-constrained agents perform lightweight local inference and rely on resource-rich edge nodes for periodic model synchronization, ensuring robust channel generalization across unseen propagation environments.
}}

\begin{figure}[t]
	\vspace{-0.33cm}
	\centering

	\subfloat[Average AoI versus transmit power.]{
		\includegraphics[width=0.94\linewidth]{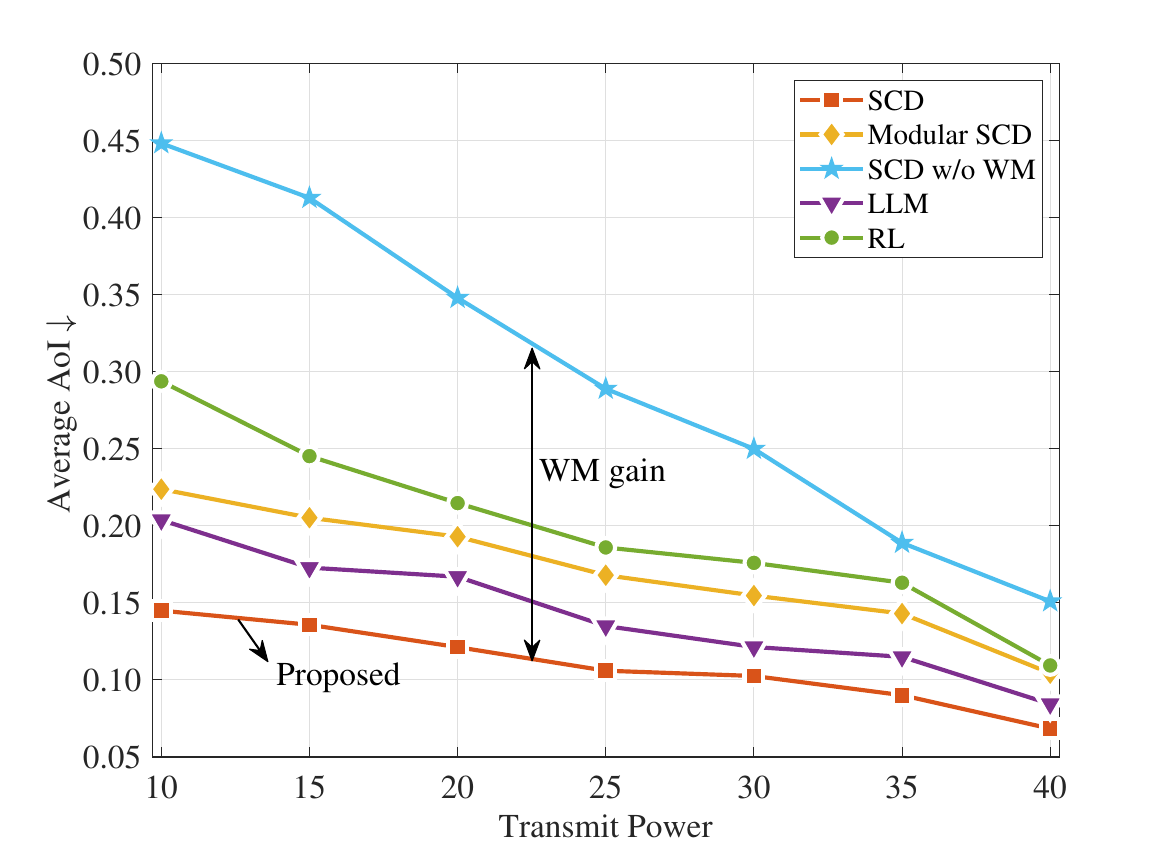}
		\label{fig:AoI_P}
	}
	\hfill

	\subfloat[Target tracking error under different SNR conditions.]{
		\includegraphics[width=0.94\linewidth]{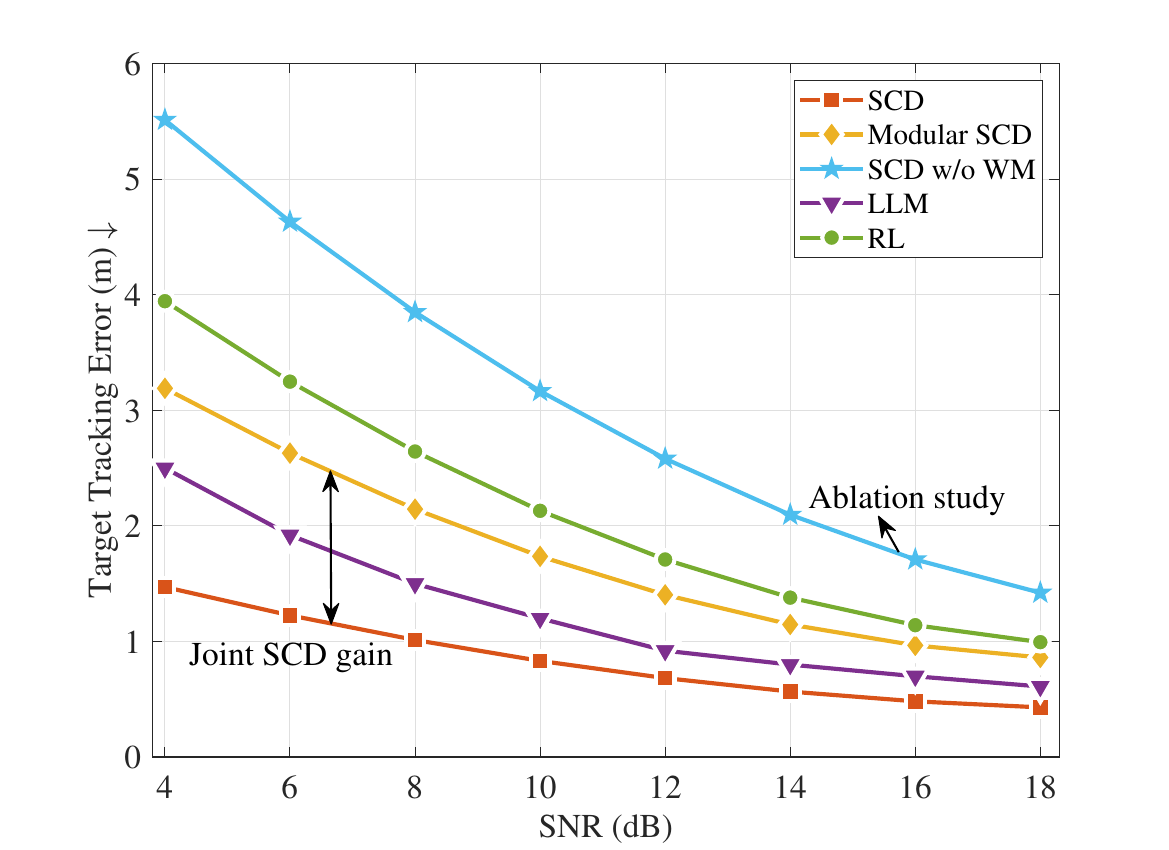}
		\label{fig:RMSE_SNR}
	}
	
	\caption{Performance comparison under different system conditions.}
	\label{fig:AoI_RMSE}
	
	\vspace{-0.4cm}
\end{figure}

\vspace{-0.2cm}
\section{Case Study}
To evaluate the proposed world model-empowered SCD framework, we conduct comprehensive simulations and analyze the results.

\vspace{-0.2cm}
\subsection{Experimental Setups}
\subsubsection{Parameters and Training Details}
For our experimental scenario, we consider multi-UAV cooperative surveillance, where a swarm of $N=5$ UAVs collaboratively tracks multiple dynamic ground targets under limited wireless resources.
Each UAV is equipped with a lightweight visual sensing module, and semantic features are extracted via a compact variational auto-encoder (VAE) to fit the world model's latent space $z[t]$.
Instead of transmitting raw images, the UAVs share compressed semantic representations by wireless links subject to Rayleigh block fading. The system bandwidth is set to 10 MHz, and the transmission power ranges from 10 to 40 dBm. 
The world model is trained using a recurrent state-space model (RSSM) architecture to handle temporal dependencies.
The decision layer adopts a proximal policy optimization (PPO)-based deep RL agent.

\subsubsection{Comparison Schemes}
For a comprehensive performance analysis, we compare the proposed framework with several representative benchmark schemes.
\begin{itemize}
	\item {Modular SCD: Sensing, communication, and decision are optimized independently.} 
		%The sensing module acquires environmental observations, the communication module performs conventional resource allocation, and the decision module determines control actions using the received data.
	\item {SCD w/o WM: This scheme adopts the same configuration as SCD but lacks predictive latent modeling and counterfactual reasoning given by the world model. }
		%Consequently, the system relies solely on instantaneous observations for decision-making without maintaining an internal latent representation of the environment.
	\item {LLM: This scheme leverages a typical large language model for multi-agent coordination by transmitting voluminous tokenized states.} 

	\item {RL: This baseline represents a conventional deep reinforcement learning approach in which the agent directly learns a policy from raw observations.} 
		%The RL agent maps instantaneous observations to actions without constructing an internal environmental model or performing predictive simulations. 
\end{itemize}

\subsection{Experimental Results}
\subsubsection{Information Freshness Performance}
To quantify the proposed framework's ability to maintain timely situational awareness, we investigate the relationship between average AoI and transmission power across different schemes.
As shown in  Fig.~\ref{fig:AoI_RMSE}{\red{(a)}}, the proposed SCD framework achieves a significantly lower average AoI compared to the baseline schemes. By leveraging world model’s predictive capability, agents can predict future state uncertainty and proactively trigger sensing updates. 
Under the experiment conditions, the world model-empowered SCD approach reduces the AoI by approximately 30\% compared with SCD w/o WM, ensuring that the decision layer operates on the most relevant environmental conditions with fresh information.

\subsubsection{Target Tracking Error Performance}
To evaluate the system robustness in highly dynamic and partially observable environments, we simulate a cooperative multi-UAV ground target tracking mission under varying channel conditions
We adopt the root mean square error (RMSE) of the estimated target trajectory as the primary performance metric. As shown in Fig. ~\ref{fig:AoI_RMSE}{\red{(b)}}, 
the proposed SCD framework consistently outperforms the baseline methods across the entire SNR regime. 
At low-SNR condition of 4 dB, the Edge LLM baseline exhibits the most severe performance degradation, yielding a tracking error of 5.20 m. This vulnerability is primarily attributed to the massive communication overhead inherent in tokenized state representations.
While the Deep RL approach avoids the heavy payload bottleneck of the LLM, SCD still has a tracking error gain of 3.60 m at 4 dB.
This gain stems directly from the world model's predictive latent rollout capability instead of passively relying on continuous real-time data reception.

%\subsubsection{MS-SSIM and LPIPS Performance}
%We conduct the performance comparison of different schemes under varying SNR levels in terms of multi-scale structural similarity index (MS-SSIM) and learned perceptual image patch similarity (LPIPS) metrics. The proposed SCD scheme consistently achieves the best reconstruction performance across all SNR regimes. For instance, at 16 dB, SCD achieves an MS-SSIM value of 0.957, which corresponds to approximately 3.9\%, 7.3\%, and 11.8\% improvements over LLM, RL, and SCD w/o WM, respectively. Similar gains can also be observed at lower SNR levels, highlighting the task effectiveness of the proposed SCD framework.

The superiority of the proposed scheme stems from three fundamental mechanisms: i) The world model rolls out future state trajectories in latent space to assess information validity over the decision horizon, triggering transmissions only when predicted deviation exceeds a threshold, thereby reducing redundant updates.  
ii) Unlike other baselines  that decouple power allocation and update scheduling, the proposed framework jointly optimizes both, prioritizing high-impact sensing tasks under power constraints to mitigate AoI escalation.
iii) Through latent simulation of multiple channel realizations, the model evaluates actions before execution, reducing transmission failures and retransmissions under fading conditions.

%\begin{table}[htbp]
%	\centering
%	\caption{Performance Comparison of MS-SSIM and LPIPS under Different SNR levels.}
%	\label{tab:performance_comparison}
%	\begin{tabular}{@{}p{1.2cm}lcccc@{}}
%		\toprule
%		& SNR(dB) & 4 & 8 & 12 & 16 \\ \midrule
%		\multirow{5}{*}{MS-SSIM} 
%		& SCD & 0.936 & 0.942 & 0.950 & {0.957} \\ \cmidrule(l){2-6} 
%		& LLM & 0.891 & 0.905 & 0.917 & 0.921 \\ \cmidrule(l){2-6} 
%		& RL& 0.850 & 0.871 & 0.884 & 0.892 \\ \cmidrule(l){2-6} 
%		& SCD w/o WM  & 0.625 & 0.689 & 0.718 & 0.762 \\ \midrule
%		\multirow{5}{*}{\quad LPIPS} 
%		& SCD & 0.132 & 0.127 & 0.121 & 0.115 \\ \cmidrule(l){2-6} 
%		& LLM & 0.084 & {0.072} & {0.069} & {0.063} \\ \cmidrule(l){2-6} 
%		& RL & 0.071 & {0.069} & {0.060} & {0.054} \\ \cmidrule(l){2-6} 
%		& SCD w/o WM & {0.063} & 0.059 & 0.051 & 0.047 \\ \bottomrule
%	\end{tabular}
%	\label{table}
%\end{table}
%
%\vspace{-0.2cm}
\section{Open Challenges and Research Directions}
Despite the promising potential of world model–empowered integrated sensing, communication, and decision, several fundamental challenges remain before such frameworks can be widely deployed in real-world complex unmanned systems.
	
\textit{1) Scalable and Stable World Model Learning in Large-Scale Heterogeneous Systems:}
As the number of heterogeneous agents (e.g., satellites, UAVs, and ground robots) increases, maintaining scalable and stable world model training and inference becomes challenging due to rapidly growing state and action spaces. Future work should explore distributed modeling, hierarchical abstraction, and lightweight representations to ensure scalability and coordination efficiency.

\textit{2) Cross-Layer Co-Design under Extreme Dynamics and Uncertainty:}
Highly dynamic environments with mobility, blockage, and intermittent connectivity pose challenges to cross-layer optimization. Small prediction errors in channel and AoI estimation may significantly affect high-level decision-making. Robust and uncertainty-aware modeling is essential for reliable performance under extreme conditions.

\textit{3) Trustworthiness, Interpretability, and Safety Assurance:}
For safety critical missions, decision transparency and reliability are crucial. Future research should explore explainable decision mechanisms, safety-constrained learning, and fault-tolerant knowledge updating to ensure trustworthy operation of complex unmanned systems.

\vspace{-0.1cm}
	\section{Conclusion}
In this article, we investigated the integration of sensing, communication, and decision for complex unmanned communication networks. 
To address the limitations of conventional modular architectures, we developed a world model–empowered joint SCD framework that established a closed-loop interaction among time-sensitive sensing, wireless communication, and intelligent decision. 
Within this framework, a time-sensitive AoI-driven sensing mechanism was introduced to regulate information freshness and reduce redundant sensing overhead. 
Furthermore, a predictive world model was constructed to capture environmental dynamics, wireless channel variations, and agent mobility within a unified latent representation, enabling proactive communication scheduling and cooperative decision evaluation. 
To facilitate large-scale heterogeneous coordination, a multi-granularity knowledge graph was  incorporated to organize cross-population relationships among agents. 
{{Future research will focus on scalable distributed learning and complex, large-scale experimental validations to thoroughly explore the practical deployment limits of the proposed space-air-ground integrated framework.}}


\begin{thebibliography}{1}
\bibitem{plan_2024}		
R. Panowicz and W. Stecz, 
``Robust Optimization Models for Planning Drone Swarm Missions,'' 
\emph{Drones}, vol. 8, no. 10, pp. 572, Oct. 2024.

\bibitem{ding2025understanding}
J. Ding \emph{et al.}, ``Understanding World or Predicting Future? A Comprehensive Survey of World Models,'' \emph{ACM Comput. Surv.}, vol. 58, no. 3, pp. 1--38, Mar. 2025.

\bibitem{MBRL}
M. Okada and T. Taniguchi, ``Dreaming: Model-based Reinforcement Learning by Latent Imagination Without Reconstruction,'' in \emph{Proc. IEEE Int. Conf. Robot. Autom. (ICRA)}, Xi'an, China, May 2021, pp. 4209--4215.


\bibitem{hafner2019dream}
D. Hafner \emph{et al.}, ``Dream to Control: Learning Behaviors by Latent Imagination,'' in \emph{Proc. Int. Conf. Learn. Represent. (ICLR)}, Addis Ababa, Ethiopia, Apr. 2020.

\bibitem{wang2025dmwm}
L. Wang \emph{et al.}, ``DMWM: Dual-Mind World Model with Long-Term Imagination,'' in \emph{Proc. Adv. Neural Inf. Process. Syst. (NeurIPS)}, San Diego, CA, USA, Dec. 2025.

\bibitem{huang2026pointworld}
W. Huang \emph{et al.}, ``PointWorld: Scaling 3D World Models for In-The-Wild Robotic Manipulation,'' \emph{arXiv preprint arXiv:2601.03782}, Jan. 2026.

\bibitem{liPuzzleItOut2026}
X. Li \emph{et al.}, ``Puzzle it Out: Local-to-Global World Model for Offline Multi-Agent Reinforcement Learning,'' \emph{arXiv preprint arXiv:2601.07463}, Feb. 2026.

\bibitem{TDMPC}
N. Hansen \emph{et al.}, ``TD-MPC2: Scalable, Robust World Models for Continuous Control,'' in \emph{Proc. Int. Conf. Learn. Represent. (ICLR)}, Vienna, Austria, May 2024.



\bibitem{robot}
P. Wu \emph{et al.}, 
``DayDreamer: World Models for Physical Robot Learning,'' in \emph{Proc. Conf. Robot Learn. (CoRL)}, Nov. 2023, pp. 2226--2240.

\bibitem{Driving}
D. Wu \emph{et al.}, ``YOLOP: You Only Look Once for Panoptic Driving Perception,'' 
\emph{Mach. Intell. Res.}, vol. 19, no. 6, pp. 550--562, Nov. 2022.



\bibitem{Driving_CVPR}
Z. Zhou \emph{et al.}, ``Query-Centric Trajectory Prediction,''  in \emph{Proc. IEEE/CVF Conf. Comput. Vis. Pattern Recognit. (CVPR)}, Jun. 2023, pp. 17863--17873.


\bibitem{wang2025world}
L. Wang \emph{et al.}, ``World Model-Based Learning for Long-Term Age of Information Minimization in Vehicular Networks,'' \emph{arXiv preprint arXiv:2505.01712}, Aug. 2025.

\bibitem{AoI}
W. Xu \emph{et al.}, ``Concept and Key Technologies of Time-Sensitive Age of 
Information in Mega Constellation Networks,'' \emph{J. Space Technol. Eng.}, vol. 2, no. 4, pp. 32--40, Feb. 2025.

\bibitem{AirScape}
B. Zhao \emph{et al.}, 
``AirScape: An Aerial Generative World Model With Motion Controllability,' 
in \emph{Proc. ACM Int. Conf. Multimedia (MM)}, Dublin, Ireland, Oct. 2025, pp. 12519--12528.

\bibitem{KG}
C. Peng \emph{et al.}, ``Knowledge Graphs: Opportunities and Challenges,'' \emph{Artif. Intell. Rev.}, vol. 56, no. 10, pp. 13071--13102, Oct. 2023.
	\end{thebibliography}
\end{document}